\documentclass[]{AO4ELT}  

\usepackage{microtype}
\usepackage{biblatex}
\usepackage{amsmath,amsfonts,amssymb}
\usepackage{graphicx}
\usepackage{pst-all} 
\usepackage[colorlinks=true, allcolors=blue]{hyperref}
\makeatletter         
\def\@maketitle{
\includegraphics[width = 170mm]{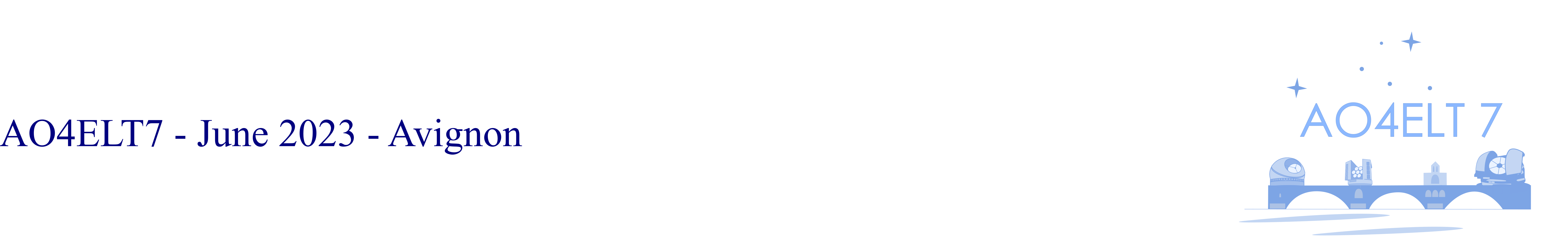}\\[8ex]
\begin{center}
{\Huge \bfseries \sffamily \@title }\\[4ex] 
{\Large  \@author}\\[4ex] 
\@date
\end{center}}

\usepackage[sc, hang]{caption}
\usepackage{graphics}
\usepackage{makeidx}
\usepackage[pass]{geometry}
\usepackage{booktabs}
\usepackage{marvosym}
\usepackage[hang]{subfigure}
\usepackage{rotating}
\usepackage{csquotes}
\usepackage{url}
\usepackage{multicol}
\usepackage{multirow}
\usepackage{overpic}
\usepackage{commath}
\usepackage{bm}
\usepackage{rotating}
\usepackage{psfrag}
\usepackage{parallel}
\usepackage{fancyhdr}

\usepackage[english]{babel} 
\usepackage{eso-pic,graphicx}
\usepackage{enumerate}
\usepackage{psfrag}
\usepackage{booktabs}

\usepackage{color, colortbl}
\definecolor{LightGray}{gray}{0.9}
\usepackage{diagbox}
\usepackage{array}
\newcolumntype{C}[1]{>{\centering\let\newline\\\arraybackslash\hspace{0pt}}m{#1}}

\title{NGSs acquisition in MORFEO}

\author[a,c]{Guido Agapito}
\author[a,c]{Lorenzo Busoni}
\author[a,c]{Cédric Plantet}
\author[a,c]{Giulia Carlà}
\author[a,c]{Marco Bonaglia}
\author[b]{Paolo Ciliegi}
\affil[a]{INAF -- Osservatorio Astrofisico di Arcetri, Largo E. Fermi 5, 50125, Firenze, Italy}
\affil[b]{INAF -- Osservatorio di Astrofisica e Scienza dello Spazio Bologna, Via Piero Gobetti 93/3, 40129 Bologna, Italy}
\affil[c]{ADaptive Optics National laboratory in Italy (ADONI)}

\authorinfo{Further author information: (Send correspondence to G.A.)\\G.A.: E-mail: guido.agapito@inaf.it}

\begin{document} 
\maketitle

\begin{abstract}
MORFEO (Multi-conjugate adaptive Optics Relay For ELT Observation) is the future multi-conjugate adaptive optics system for the ESO ELT that will feed the instrument MICADO (Multi-AO Imaging Camera for Deep Observations). It will use the 6 laser guide stars to give a uniform correction on a field-of-view of approximately 60arcsec of diameter. Tip, tilt and slow focus measurement will be done on up to three natural guide stars that could be really faint to maximize sky coverage. The current baseline is to use the reference wavefront sensor in the visible to acquire the star and center it on the low order wavefront sensor that has a much smaller field-of-view. In this work we study this problem focusing on the estimation error of the tilt from the reference wavefront sensor as a function of star magnitude and atmospheric conditions.
\end{abstract}

\keywords{extremely large telescope, multi conjugate adaptive optics, natural guide star, star acquisition, wavefront sensing, simulations}


\section{Introduction}
\label{sect:intro}  

MORFEO (Multi-conjugate adaptive Optics Relay For ELT Observation, formerly known as MAORY) will provide multi conjugate adaptive optics correction to MICADO~\cite{2021Msngr.182...17D}.
It will grant a uniform performance thanks to 6 Laser Guide Stars (LGSs), 3 Natural Guide Stars (NGSs) and 3 Deformable Mirrors (DMs): an average K band SR greater than 30\% on half of the sky in median atmospheric condition and greater than 50\% on the best quartile of atmospheric conditions.
Further details are provided in Refs. \cite{2021Msngr.182...13C,2022SPIE12185E..14C,2022SPIE12185E..4RB,BusoniAO4ELT7}.

MORFEO relies on the tilt and focus measurements from up to three NGSs to achieve the desired performance over half of the sky.
This implies that it will work with very faint stars, in fact, currently, the limiting magnitude is H=21, that is about 3000 detected photons per second in H band (the total throughput of the low order wavefront sensor - LO WFS - is about 0.3).
Acquiring such faint stars is a key step in the operation of MORFEO and other Multi-Conjugate Adaptive Optics (MCAO) systems for the next generation of telescopes such as NFIRAOS/IRIS\cite{Andersen2019_2}.
In MORFEO the Field-of-View (FoV) of the LO WFS is 1.9 arcsec and we expect (with the current limited knowledge on the telescope pointing) this value to be not enough to acquire directly the NGS.
So, most of the times the acquisition will be done at first with the reference WFS\footnote{note that we do not consider the option to use MICADO for acquisition because (1) the NGSs will be outside the science FoV to avoid vignetting and (2) to avoid possible persistence issue on the science detector.} (RWFS) thanks to its larger FoV of about 4.6 arcsec.
This WFS is a visible 8$\times$8 Shack-Hartmann (SH) sensor used as tomographic truth sensor with the main objective in reducing the bias due to non common path aberration between science path and LGS WFS path and, to a lesser extent, to truncation effects.
As mentioned above, it has a larger FoV of about 4.6 arcsec and it is pointing to the same star of the LO WFS: in fact, a dichroic is splitting the light coming from the same star around 1$\mu$m between LO WFS and RWFS (see Ref. \cite{2022SPIE12185E..4OB}, LO WFS is working in H band, RWFS is working in R+I bands).

This article is structured in a couple of sections: Sec. \ref{sec:acqSeq} presents the acquisition sequence and Sec. \ref{sec:acqtime} shows the estimation of the time required to acquire the NGS.

\section{NGS acquisition sequence}\label{sec:acqSeq}

\begin{figure}[h]
    \centering
    \includegraphics[width=0.5\linewidth]{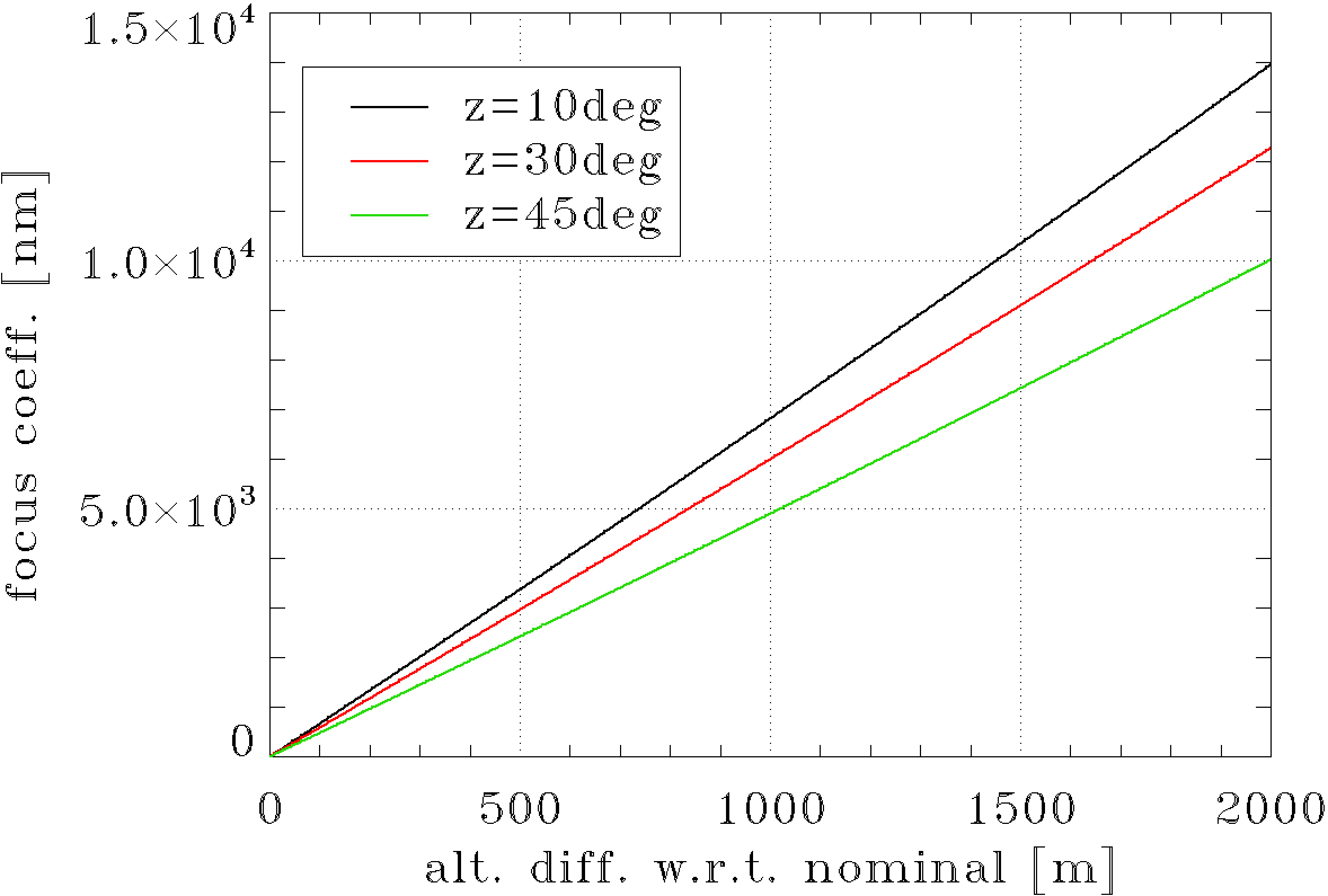}
    \caption{Focus coefficient as a function of the difference between sodium average altitude and nominal value at zenith (90km).}
    \label{fig:focusErrorNaAlt}
\end{figure}

\begin{figure}[h]
    \centering
    \includegraphics[width=0.55\linewidth]{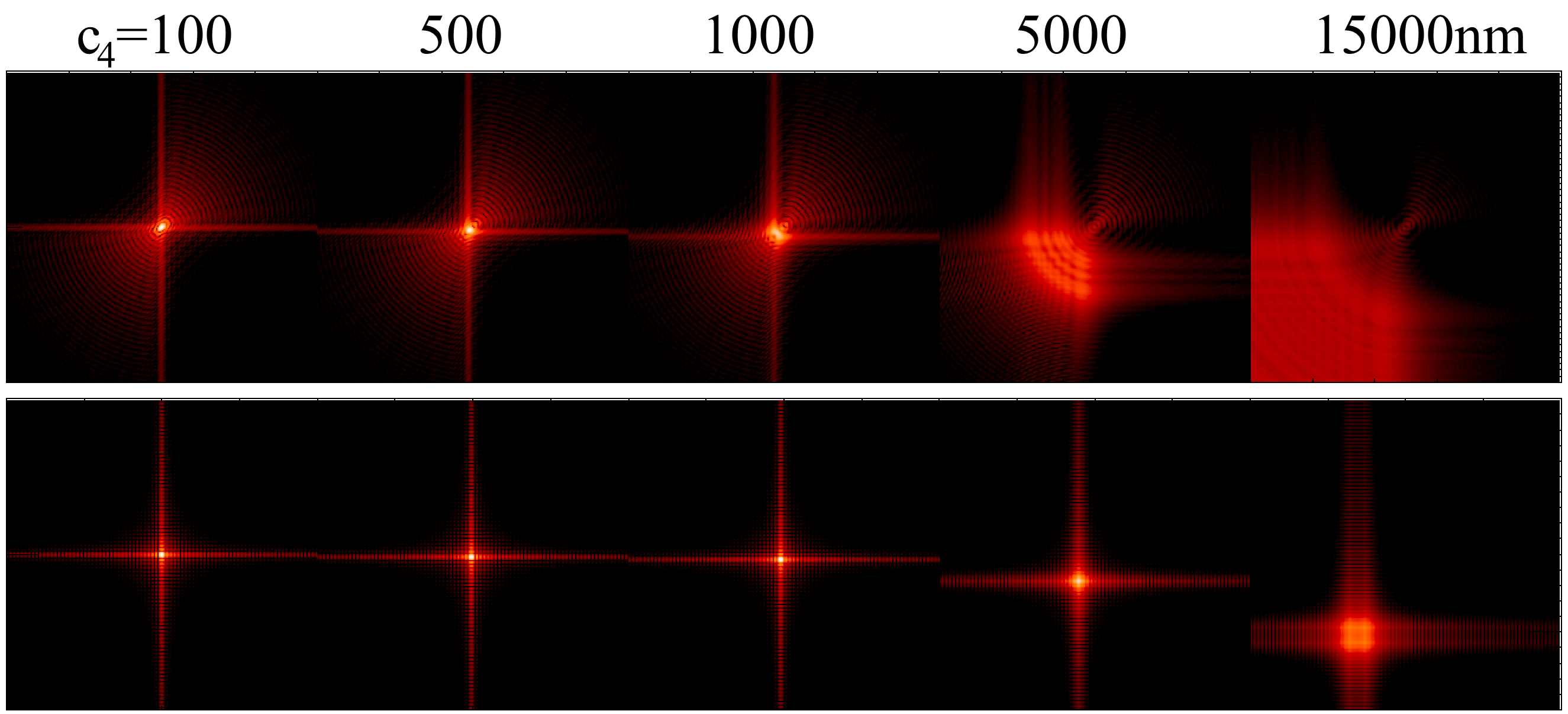}
    \caption{Effect on a sub-aperture of the (top) 2$\times$2 LO SH and (bottom) 8$\times$8 reference SH of a focus with a coefficient ranging from 100 to 15000nm. In case of reference SH the considered sub-aperture is at the edge of the pupil. Note that no other aberration is present in the wavefront sensed by the two sensors.}
    \label{fig:focusEffectSa}
\end{figure}

The NGS acquisition will be preceded by the closure of the LGS loop.
This will give the correction of the so-called high order modes: all modes except tip and tilt will be corrected.
This correction will be biased on focus due to sodium layer average altitude and on a few low order modes due to the elongation/truncation effects.
The bias due to elongation/truncation will be small thanks to the approach described in Refs. \cite{oberti:hal-02614170}, \cite{2020SPIE11448E..2SA} and \cite{2022JATIS...8b1514F}, but the focus error can be large, because we can expect the sodium layer to be out of a few hundreds of meters from the standard value of 90 km (see Ref. \cite{2014A&A...565A.102P}, Fig. 5) and the propagation to focus coefficient on a 39-m telescope is huge as can be seen in Fig. \ref{fig:focusErrorNaAlt}. 
This focus error can greatly reduce the sensing capability, in particular on the LO WFS as can be seen in Fig. \ref{fig:focusEffectSa}.
So, a method that roughly estimate the sodium layer altitude is required to improve this first step allowing for a shorter acquisition. 
This could be done by using a reference focus value that compensates for the nominal focus introduced by the optics and excluding LGS focus compensation or by moving the LGS focus stage to null the average focus measured by the LGS WFSs w.r.t. the reference focus value.
\begin{figure}[h]
    \centering
    \includegraphics[width=0.65\linewidth]{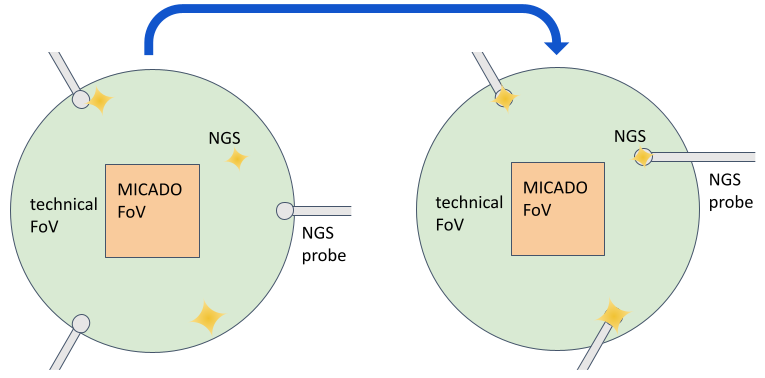}
    \caption{Scheme of the first step of the NGS acquisition: probes are moved from the initial position (left part) to the nominal guide stars position (right part). Technical and MICADO/science FoV are shown in green and orange colors respectively.}
    \label{fig:NGS_acquisition_probes}
\end{figure}
\begin{figure}[h]
    \centering
    \includegraphics[width=0.65\linewidth]{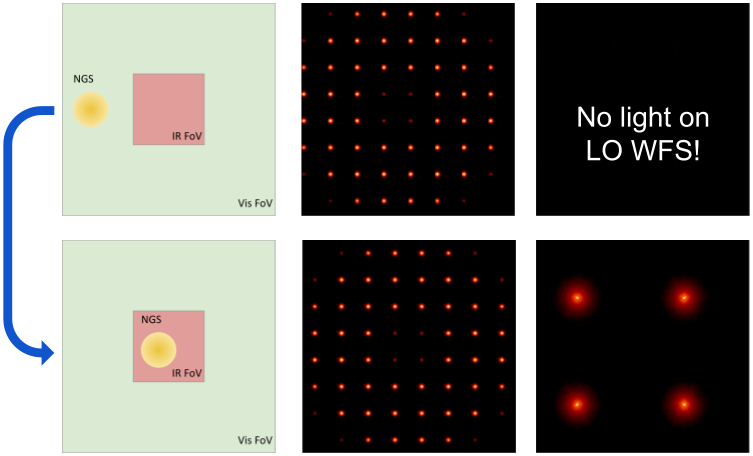}
    \caption{Scheme of the NGS acquistion step on the RWFS when the initial position of the guide star is outside the LO WFS FoV but inside the RWFS one. Left, guide star position on the FoVs of the infrared (LO WFS, red color) and visible (RWFS, green color) channels, center, the visible (RWFS) detector image, right, the infrared (LO WFS) detector image. On top the star is in the RWFS FoV, but not in the LO WFS one, then, bottom, after the acquisition on the RWFS the star falls in the LO WFS FoV.}
    \label{fig:NGS_acquistion_REF_and_LO}
\end{figure}
\begin{figure}[h]
    \centering
    \includegraphics[width=0.75\linewidth]{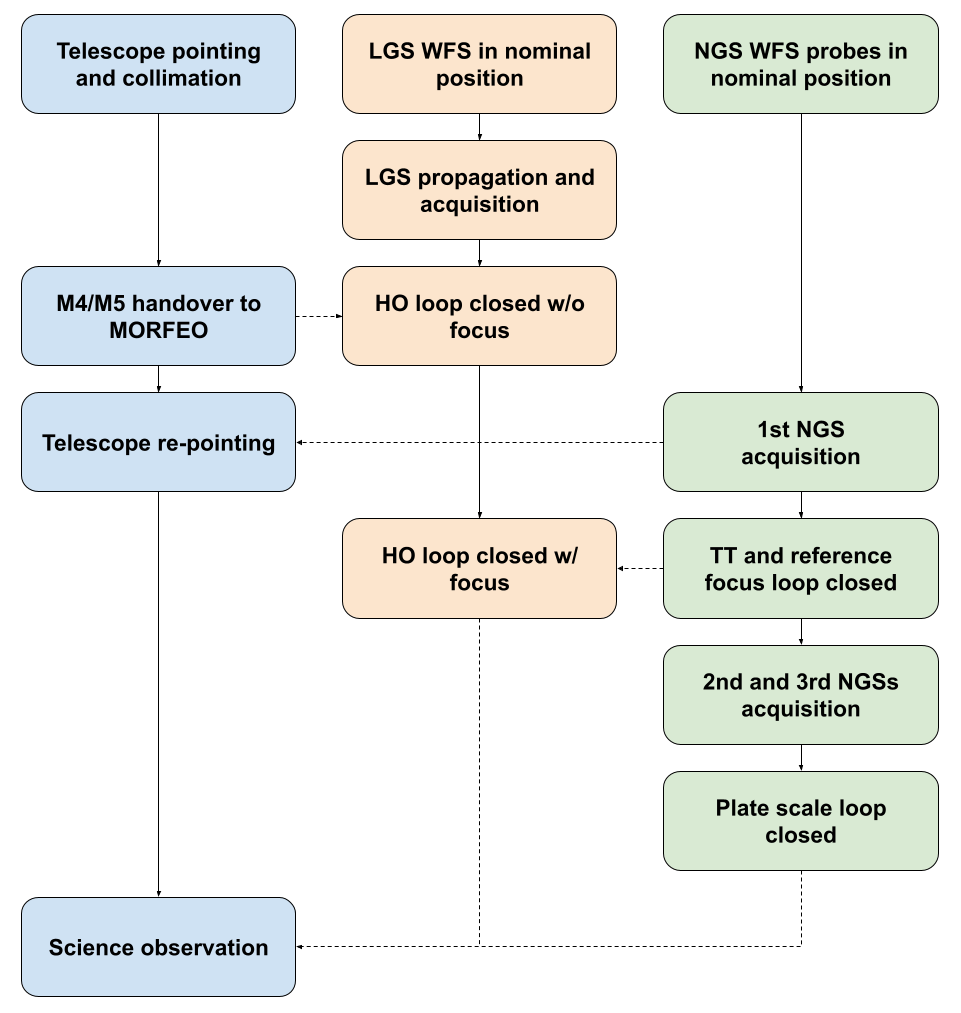}
    \caption{MORFEO acquisition sequence diagram (\emph{happy path} scenario). Time flows from top to bottom. Blue color indicate telescope/instrument, orange LGS WFS and green NGS WFS.}
    \label{fig:MORFEO_acquistion_diagram}
\end{figure}

In parallel to the telescope pointing, LGS propagation and acquisition, the three NGS probes are positioned in the nominal NGS positions.
Then, the NGS acquisition starts from the brightest star when the LGS loop is closed.
In this first step tilt and focus are measured on the reference WFS to center the star on the LO WFS (by re-pointing the telescope) and to correct for the focus bias due to the unknown sodium layer altitude.
After the first NGS is acquired and overall tilt and focus correction is active the acquisition of the two other NGSs is done in parallel to minimize the overall acquisition time.
So, the summary of the acquisition (also reported in Fig. \ref{fig:MORFEO_acquistion_diagram}) considering an \emph{happy path} scenario is:
\begin{enumerate}
    \item Position LGS WFSs (a rotation function of zenith angle) and NGS probes (see Fig. \ref{fig:NGS_acquisition_probes}, note that field is derotated by MICADO rotator).
    \item Acquire LGSs.
    \item Close LGS loop without focus (LGS WFS stage moved to obtain an average zero focus signal on LGS WFSs).
    \item Acquire brightest NGS on reference WFS (see Fig. \ref{fig:NGS_acquistion_REF_and_LO}).
    \item Re-point telescope and correct average focus.
    \item Acquire brightest NGS on LO WFS.
    \item Close tip, tilt and focus loops (LGS WFS stage in closed loop on the LO WFS focus signal and M4 in closed loop on LGS WFS focus signal).
    \item Acquire the other NGSs on reference WFSs.
    \item Move probes (to avoid requiring high strokes on DM due to platescale correction).
    \item Acquire the other NGSs on LO WFSs.
    \item Close plate scale loop.
\end{enumerate}
Note that when an offset larger than LO WFS FoV (that means the probes position must be changed) is performed part of this sequence must be replicated.
When the probes are moved to a new position, at least acquisition of NGSs on the LO WFSs must be done and if no star is present in the LO WFS FoV also the acquisition of NGSs on the reference WFSs.
In the case of offset we expect that possible errors are common to all the three probes and can be corrected by a tip and tilt correction.
Note that a description of the NGS acquisition from the instrument control system software point of view is presented in Ref. \cite{2022SPIE12189E..1VS}. 

The estimation of the NGS acquisition time is presented in the next section. 

\section{NGS acquisition time}\label{sec:acqtime}

The NGS acquisition time will be dominated by the time required by the reference sensors to measure the position of the NGS with enough accuracy to bring the star in the FoV of the LO sensors.
This measurement will be done exposing the detector for a given time and calculating the centroid of the spots in the valid sub-apertures of the reference sensor, so as to estimate the position of the star.
The centroid will be computed with a center-of-gravity with a threshold (TCoG) algorithm because the unknown initial position of the star does not allow to use windows or weighting maps.
The threshold is used to reduce errors induced by read-out noise, dark current and sky background.
\begin{table}[ht]
\caption{Summary of the MORFEO parameters used in simulation.}
\label{Tab:params}
\begin{center}
\begin{small}
	\begin{tabular}{|l|c|}
		\hline
		\textbf{Parameter} & \textbf{value}\\
		\hline
		Telescope diameter &  38.5m\\
		Central obstruction & 11.0m\\
		Pupil mask & M1 pupil with spiders\\
            $\epsilon$ (line-of-sight) & 0.71 -- 1.09arcsec\\
		$L_0$ & 25m\\
		RWFS off-axis angle & 30 -- 80arcsec\\
  		RWFS no SA & 8$\times$8\\
		RWFS FoV & 4.6arcsec\\
		RWFS pixel pitch & 153mas\\
            RWFS bandwidth & 600 -- 1000nm\\
            RWFS full throughput & 0.154\\
            RWFS zero magnitude & $1.82 \cdot 10^{-8}$ $\mathrm{J/s/m^2/\mu m}$\\
		RWFS sky background (no moon) & 43 $\mathrm{e^-/m^2/arcsec^2/s}$\\
		RWFS sky background (moon )& 453 $\mathrm{e^-/m^2/arcsec^2/s}$ (moon at 20deg)\\
		RWFS detector RON & 0.2$\mathrm{e^-/pixel/frame}$\\
            RWFS excess noise factor & 2 (variance)\\
		RWFS detector dark current & 0.58$\mathrm{e^-/pixel/s}$\\
		RWFS Centroiding algorithm & CoG with threshold\\
		\hline
		\multicolumn{2}{c}{\scriptsize Note: $\epsilon$ is seeing, RWFS is Reference Wavefront Sensor, SA is Sub-Aperture, }\\
		\multicolumn{2}{c}{\scriptsize FoV is Field-of-View, RON is Read-Out Noise and CoG is Center of Gravity.}\\
        \multicolumn{2}{c}{\scriptsize Some values are not the same as in Ref. \cite{BusoniAO4ELT7}, as they were updated after this analysis.}
	\end{tabular}
\end{small}
\end{center}
\end{table}
\begin{table}
\caption{Integration time of the RWFS, T$_{int}$, required to get an estimation error of the NGS position that is less than 0.35arcsec RMS (about $1/4$ of the LO FoV). The integration time is a function of seeing, $\epsilon$, NGS off-axis distance, presence of moon (full moon at 20deg from target) and R magnitude.}
\label{Tab:intTime}
\begin{small}
\begin{center}
	\begin{tabular}{|c|c|c|c|c|}
		\hline
	  \textbf{$\epsilon$ [asec]} & \textbf{NGS off-axis} &  \textbf{Moon} &  \multirow{2}{*}{\textbf{R magn.}} & \textbf{req.} \\
   \textbf{(line of sight)} & \textbf{dist. [asec]} & \textbf{in sky} &  & \textbf{T$_{int}$ [s]} \\
		\hline
        \multirow{3}{*}{0.71} & \multirow{3}{*}{30} & \multirow{3}{*}{No} & 21 & 1.0 \\
         & & & 22 & 2.0 \\
         & & & 23 & 7.5 \\
		\hline
        \multirow{3}{*}{1.09} & \multirow{3}{*}{30} & \multirow{3}{*}{No} & 21 & 1.0 \\
         & & & 22 & 2.0 \\
         & & & 23 & 10.0 \\
		\hline
        \multirow{3}{*}{0.71} & \multirow{3}{*}{30} & \multirow{3}{*}{Yes} & 21 & 2.0 \\
         & & & 22 & 7.5 \\
         & & & 23 & 45.0 \\
		\hline
        \multirow{3}{*}{1.09} & \multirow{3}{*}{30} & \multirow{3}{*}{Yes} & 21 & 2.5 \\
         & & & 22 & 12.5 \\
         & & & 23 & 75.0 \\
		\hline
        \multirow{3}{*}{0.71} & \multirow{3}{*}{55} & \multirow{3}{*}{No} & 21 & 1.0 \\
         & & & 22 & 2.0 \\
         & & & 23 & 7.5 \\
		\hline
        \multirow{3}{*}{1.09} & \multirow{3}{*}{55} & \multirow{3}{*}{No} & 21 & 1.0 \\
         & & & 22 & 2.0 \\
         & & & 23 & 10.0 \\
		\hline
        \multirow{3}{*}{0.71} & \multirow{3}{*}{55} & \multirow{3}{*}{Yes} & 21 & 2.0 \\
         & & & 22 & 10.0 \\
         & & & 23 & 75.0 \\
		\hline
        \multirow{3}{*}{1.09} & \multirow{3}{*}{55} & \multirow{3}{*}{Yes} & 21 & 3.0 \\
         & & & 22 & 15.0 \\
         & & & 23 & 90.0 \\
		\hline
        \multirow{3}{*}{0.71} & \multirow{3}{*}{80} & \multirow{3}{*}{No} & 21 & 1.0 \\
         & & & 22 & 2.0 \\
         & & & 23 & 7.5 \\
		\hline
        \multirow{3}{*}{1.09} & \multirow{3}{*}{80} & \multirow{3}{*}{No} & 21 & 1.3 \\
         & & & 22 & 5.0 \\
         & & & 23 & 15.0 \\
		\hline
        \multirow{3}{*}{0.71} & \multirow{3}{*}{80} & \multirow{3}{*}{Yes} & 21 & 2.0 \\
         & & & 22 & 12.0 \\
         & & & 23 & 75.0 \\
		\hline
        \multirow{3}{*}{1.09} & \multirow{3}{*}{80} & \multirow{3}{*}{Yes} & 21 & 5.0 \\
         & & & 22 & 20.0 \\
         & & & 23 & 120.0 \\
		\hline
	\end{tabular}
\end{center}
\end{small}
\end{table}
\begin{figure}[h]
    \centering
    \includegraphics[width=0.55\linewidth]{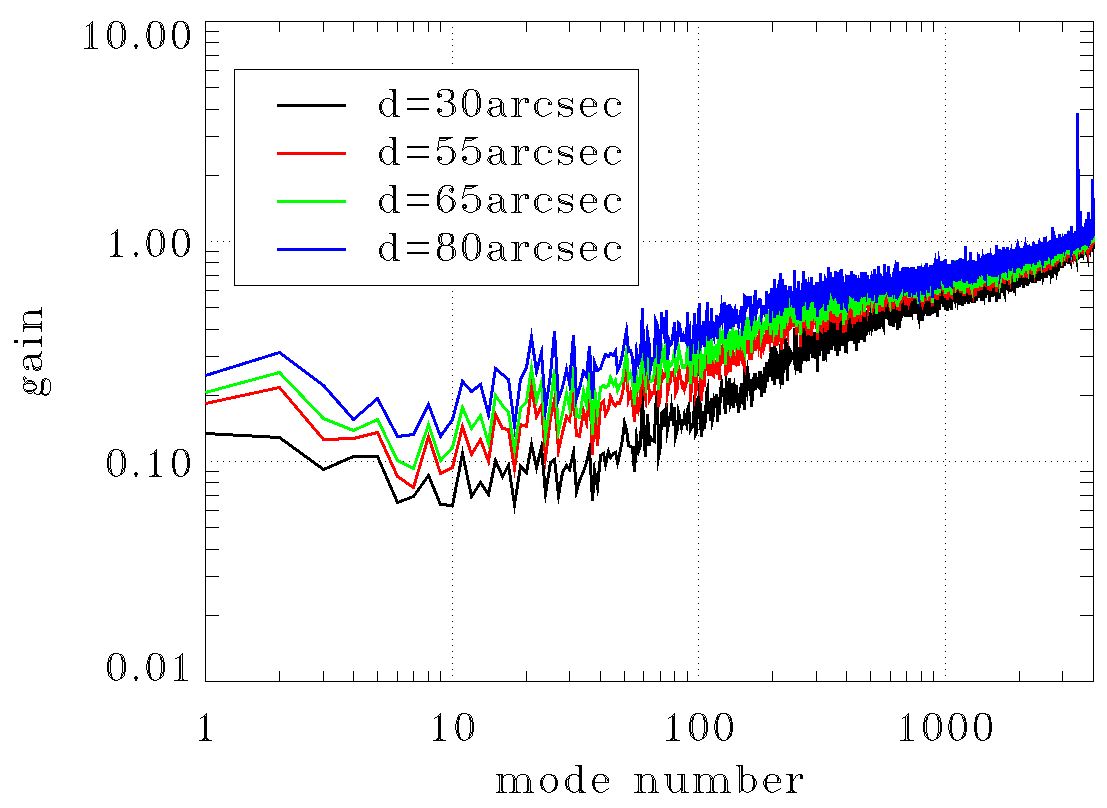}
    \caption{Modal correction coefficients, that are the ratio between residual and turbulence mode-by-mode, for different off-axis angles. Mode 1 is tip.}
    \label{fig:modal_correction_coefficients}
\end{figure}
\begin{figure}[h]
    \centering
    \includegraphics[width=0.65\linewidth]{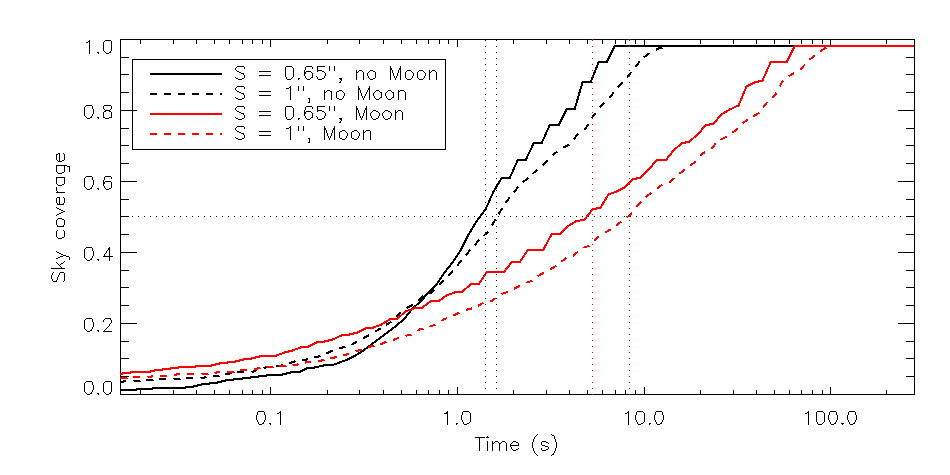}
    \caption{Sky coverage as a function of integration time required on the RWFS to acquire the dimmest star of the NGS asterism. The NGS asterisms are selected as described in Ref. \cite{2022JATIS...8b1509P} for median atmospheric condition and pointing to the south galactic pole. The limiting magnitude is R=23.}
    \label{fig:sky_cov_acquisition_time_R23}
\end{figure}

We have two ways of estimating the error associated with this TCoG:
\begin{itemize}
    \item analytical computation: fast but less accurate mainly because it considers a pessimistic case where the spot is centered on 4 pixels\footnote{in general the spot can be found in any position and when it is centered on 4 pixels the maximum flux in the sub-aperture is minimum because the light in the PSF peak is divided by 4 (considering that half of the peak size is less than 1 pixel).}.
    \item end-to-end simulation: a simplified single conjugate simulation (done in PASSATA~\cite{doi:10.1117/12.2233963}) is used to compute the error on a set of 1000 random realizations.
\end{itemize}
We then used the analytical calculation to select the initial parameters of the simulations, such as integration time and threshold level for each configuration (seeing, off-axis distance, moon, and NGS magnitude), and performed the final estimation with the end-to-end tool.
The list of parameters used in simulation are shown in Tab. \ref{Tab:params} (some parameters are not the same as Ref. \cite{BusoniAO4ELT7}, as they were updated after this analysis was conducted).
Note that the residual level is estimated by a full MOREFO end-to-end simulation as the one shown in Ref. \cite{2020SPIE11448E..2SA}: an example of the modal correction coefficients (ratio between residual and turbulence mode-by-mode) are shown in Fig. \ref{fig:modal_correction_coefficients}.

The results of the simulations are summarized in Tab. \ref{Tab:intTime} where the integration time of the RWFS required to get an estimation error of the NGS position that is less than 0.35arcsec RMS (about $1/4$ of the LO WFS FoV) is reported as a function of seeing, NGS off-axis distance, presence of moon (full moon at 20deg from target) and R magnitude.
We can see that integration times are below one second in most cases for magnitudes R$<$21 and that the main issue for the NGS acquisition on the RWFS is the moon: when it is present the integration times increases by about a factor 10.

Finally, we used the results from these simulations to estimate the sky coverage as a function of integration time required on the RWFS to acquire the dimmest star of the NGS asterism.
The NGS asterisms are selected as described in Ref. \cite{2022JATIS...8b1509P} for median atmospheric condition and pointing to the south galactic pole. 
This time should be the dominant term in the overall acquisition time of the NGSs, because the acquisition time on the brightest star of the asterism will be shorter, the two dimmest stars are acquired in parallel, and the acquisition on the LO WFS will be fast (average R-H color is 2, LO WFS sub-apertures have 16 times the area of the RWFS sub-apertures and background given by moon is negligible in H band).
Following the results of this analysis, we have decided to define as limiting magnitude R=23 because for dimmer magnitudes the acquisition time will be longer than 10s without moon and longer than 100s with moon at 20deg (this means that we no longer have a limiting magnitude specified in H band).

\section{Conclusion}

We presented the concept of the NGS acquisition in MORFEO.
This is a critical step to provide the high sky coverage required for such a system.
This concept is based on several steps and we focused on the initial acquisition with the RWFS.
We estimated the integration time required by the RWFS to move the star on the LO WFS FoV and the sky coverage as a function of this integration time.
We selected as limiting magnitude R=23: this choice was made for acquisition reason, because LO WFS in the infrared could in principle work with dimmer stars.
Moon light has a strong impact because we are using visible light in the RWFS and we know that if the initial error (pointing + plate scale) on the NGS position is smaller than about 1arcsec we can skip the acquisition on the reference WFS and greatly reduce the acquisition time, but at present we do not have such knowledge of the telescope and MORFEO that we can make this assumption on the initial error on the NGS position.
Finally, in this analysis we have neglected NGS tip and tilt jitter during the acquisition of the brightest star. This acquisition will be in general short, but nonetheless some jitter will be present if field stabilization from the ELT~\cite{2020SPIE11445E..1ET} in parallel to HO correction from LGSs will not be available.
Further work will be done when MORFEO WFS parameters are finalized after the final design review will be passed, considering also the expected tip and tilt jitter and focus error.



\printbibliography 

\end{document}